# Content Format and Quality of Experience in Virtual Reality

**Henrique Galvan Debarba · Mario Montagud · Sylvain Chagué · Javier Lajara · Ignacio Lacosta · Sergi Fernandez Langa · Caecilia Charbonnier**

**Abstract** In this paper, we investigate three forms of virtual reality content production and consumption. Namely, 360 stereoscopic video, the combination of a 3D environment with a video billboard for dynamic elements, and a full 3D rendered scene. On one hand, video based techniques facilitate the acquisition of content, but they can limit the experience of the user since the content is captured from a fixed point of view. On the other hand, 3D content allows for point of view translation, but real-time photorealistic rendering is not trivial and comes at high production and processing costs. We also compare the two extremes with an approach that combines dynamic video elements with a 3D virtual environment. We discuss the advantages and disadvantages of these systems, and present the result of a user study with 24 participants. In the study, we evaluated the quality of experience, including presence, simulation sickness and participants' assessment of content quality, of three versions of

Henrique Galvan Debarba
Artanim Foundation, Chemin du Grand-Puits 40, 1217 Meyrin, Switzerland
IT University of Copenhagen, Rued Langgaards Vej 7, 2300 Copenhagen, Denmark E-mail: hend@itu.dk

Mario Montagud
Universitat de València, Av. de Blasco Ibáñez 13, 46010 Valencia, Spain
Fundació i2CAT, Carrer del Gran Capità 2, 08034 Barcelona, Spain

Sylvain Chagué
Artanim Foundation, Chemin du Grand-Puits 40, 1217 Meyrin, Switzerland

Javier Lajara
Entropy Studio, C. Hermanos Argensola 2, 50001 Zaragoza, Spain

Ignacio Lacosta
Entropy Studio, C. Hermanos Argensola 2, 50001 Zaragoza, Spain

Sergi Fernandez Langa
Fundació i2CAT, Carrer del Gran Capità 2, 08034 Barcelona, Spain

Caecilia Charbonnier
Artanim Foundation, Chemin du Grand-Puits 40, 1217 Meyrin, Switzerland

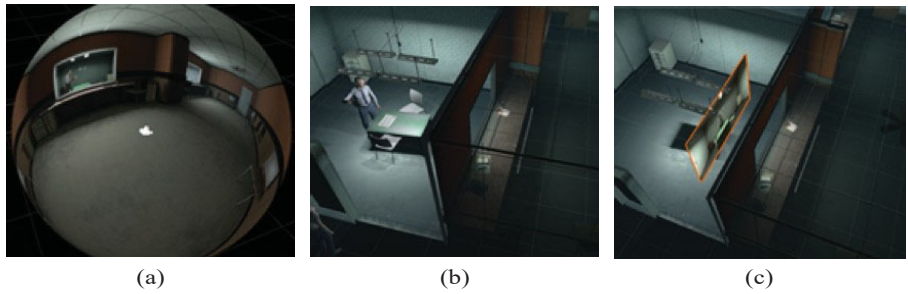

(a)                              (b)                              (c)

Fig. 1: Overview of the virtual scene in the three different content presentation formats investigated in this paper: (a) stereo VR360 video, with the video of each eye mapped into a sphere that is centered at the virtual camera position of that eye; (b) full 3D environment and cinematic; (c) 3D environment combined with video billboard cinematic, the billboard is drawn behind a window to simulate motion parallax when the user moves.

a cinematic segment with two actors. We found that, in this context, mixing video and 3D content produced the best experience.

## 1 Introduction

The recent availability of virtual reality (VR) consumer equipment has boosted the demand for content and experiences for these devices, as well as for content production technology and techniques. A range of new and affordable capture hardware, such as 360 degrees (or omnidirectional) cameras and motion tracking equipment (e.g. suits with inertial sensors), became available in the past few years. In fact, large content distribution and sharing platforms now support VR content and experiences, which can be produced and released by professional studios as well as by independent producers. Among these, we have stereoscopic 360 video streaming websites, such as Youtube, and VR games distribution services, such as Valves' Steam.

    To preserve visual consistency and achieve a high degree of visual fidelity, VR content often consists of stereoscopic 360 videos or full 3D simulations. The production pipeline and the possibilities of interaction with each form of content are different and, as a consequence, each method comes with trade-offs. For instance, 360 stereo video is commonly used to present content that is captured from a real world environment. Since it consists of video capture, the result is visually accurate and photorealistic. However, stereo 360 video does not support point of view (POV) translation since the spatial content is captured and recorded as a spherical projection from a single position in space at any given time. The lack of POV translation is a core limitation since it blocks the exploration of the space and is one of the main factors leading to

discomfort and simulation sickness in VR experiences [9]. We note that current VR headsets support position tracking and, thus, stereo 360 video content format sub-utilizes the capabilities of the hardware. On the other hand, full 3D content is normally used for interactive experiences, such as games. It recreates and stores the geometrical information and properties of the content in a 3D format, and produces 2D projections at the time of content consumption, i.e. it renders images in real-time. Thus, full 3D content can support POV translation. However, real time photorealistic rendering can only be achieved at a very high monetary and computational cost. It requires significant effort to create and animate 3D content, and to simulate the physical behavior of objects and their interactions with the light in real-time.

Beyond these two widespread VR content formats, this study explores the use of video billboards, or impostors, to merge dynamic content captured in video with 3D structural and peripheral environment information. In 3D applications, such as games, this technique is generally used to represent background elements or objects that are too far away from the point of view to be properly identified [6], a common example are the large crowds in sport games. Although the use of billboards in 3D applications is not new, in this paper we rely on billboards to represent a central — instead of peripheral — aspect of the experience. A major consequence is that the incongruent projection of the video, which does not respond correctly to POV movements, can be easily noticed by the user, as studied by Fourquet et al. [3]. Whilst being aware of these limitations, our interest is in better understanding and determining how detrimental these artefacts are to the overall quality of experience (QoE) when compared to stereo 360 video and full 3D versions of the same content.

To that end, we designed and conducted a user study to evaluate the impact of different VR content formats on the perceived user experience (QoE, immersion, simulation sickness). As having appropriate use cases and content is key for the evaluation of a novel technology/medium, we produced a professional VR content episode, with three versions resembling the conditions to be evaluated, namely: a stereo 360 version (Figure 1a); a full 3D version (Figure 1b); and a version that combines 3D environment and video content as flat billboards added to the scene (Figure 1c). We conducted a user study with 24 participants, and the results showed that, under the specifications and particularities of our content and experimental design, users often concluded that the combination of video billboards and 3D environment offered the best experience. In summary, our contributions consist of practical considerations about the process of producing VR content in different formats, discussing the intrinsic advantages and disadvantages of each format; and the evaluation of how the different content formats affect the overall user experience.

This paper is organized as follows. In Section 2, we describe the three content formats that we investigate in this paper and discuss their pros and cons. Section 3 provides an overview of the content that we created and of the production process of the three different versions of that content. Section 4 describes the design as well as the results of a user study comparing the three versions of our content. Finally, Section 5 presents a discussion and our conclusion on the topic.

## 2 Content Formats for VR

Up to date, a variety of content formats and rendering technologies for VR scenarios can be adopted, each one with different implications. The most common techniques are 360 degrees video (VR360 from now on) and full 3D environments, but hybrid solutions, like 3D scenarios with inserted video billboards for specific elements of the scene, can also provide satisfactory results. Next, an overview of their a priori main pros and cons is provided, and previous works having adopted these content formats are reviewed. The goal of the experiment presented in this work is to confirm, and gain deeper insights about, these assumptions.

### 2.1 VR360

VR360 videos represent a simple and cheap, yet effective and realistic, way to provide VR experiences. In VR360 videos, a view in every direction is recorded at the same time using an omnidirectional camera or a camera rig that captures overlapping angles simultaneously. The multiple views are then stitched together into a single, high resolution and seamless panoramic video. The camera (rig) represents the center of the omnidirectional scene, and during consumption, the users viewpoint is also placed at the center of the sphere (see Figure 2).

VR360 can be used in social interactions, Pece et al. [10] proposed a coherent representation of a meeting room with remote participants by making video inserts of the users in a 360 picture. Their solution stitches the incoming video from participants into a 360 panorama picture to anchor the participants to physical positions that are represented in the picture. A similar approach is adopted in the Social VR platform by Gunkel et al. [5].

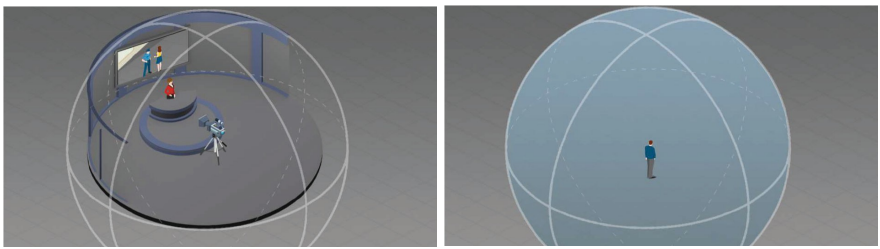

Fig. 2: Capturing and consumption viewpoint in VR360.

*Pros of VR360*

- VR360 videos are the simplest and cheapest solution in terms of VR content production, when the scenarios to be captured exist and do not need to be created as a model.
- VR360 videos provide high degree of realism, as real scenarios can be captured with high resolution and photographic quality, by using professional cameras, and virtual scenarios can be rendered at very high quality since there are no time constraints. This is especially relevant for dynamic characters.

*Cons of VR360*

- VR360 videos require a calibration of the employed cameras and the stitching of the captured images. However, existing software tools can effortlessly and successfully provide these features.
- VR360 videos are captured from a single point from where the camera (rig) is physically placed (see Figure 2). That means that the users viewpoint is static, matching the cameras position. If the user moves his/her position, then unpleasant parallax effects will soon appear, giving the feeling that the viewpoint also changes and resulting in a perceived deformation of the VR environment. Therefore, POV translation is not supported in VR360 videos, and the point of view has to be defined before recording. Although further cameras at other positions could be used, they could interfere the production and considerably increase the production efforts and costs.
- Beyond the POV translation issues, it should be noted that VR360 videos, even when using stereoscopic recordings, are flat content formats. Therefore, free navigation around the VR environment, which is commonly known as 6 Degrees of Freedom (6DoF), is not supported either.
- In the case that volumetric elements, e.g. users, need to be added in the VR experience, processing and transformation processes are necessary to properly represent them in the VR360 environment. The transformation can be done by applying a 2D mapping of the 3D volumetric data. The volumetric data will be placed in the 3D world and then projected to the 360 sphere where the rest of the video is represented (Figure 3). This feature is not considered in the presented experiment, but it is important to be kept in mind while deciding on the most appropriate format(s) when producing VR experiences.

## 2.2 Full 3D

In Full 3D, the whole VR environment, including the characters, is represented in 3D. This content typology is widely used in VR experiences. An example of a Full 3D environment can be seen in Figure 4, where the building, characters, and end-users are presented in volumetric 3D.

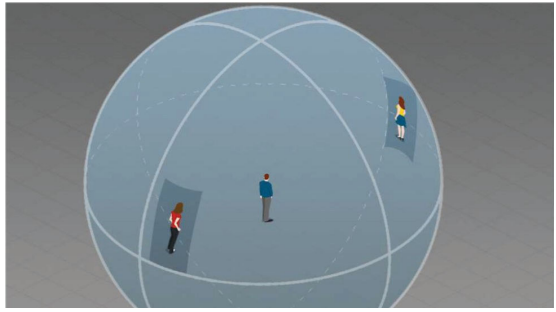

Fig. 3: Projected 3D volumetric elements on a 2D 360 degrees sphere.

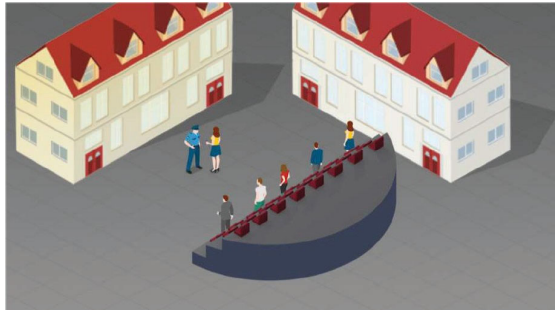

Fig. 4: Example of a Full 3D environment.

With regard to the content production, the VR environment can be 3D modelled from scratch, and also 3D scanners can be used for existing scenarios. The representation of the characters can be achieved by making use of scanning techniques to create 3D avatars, which can then be animated using Motion Capture (MoCap) techniques. The use of real-time capturing solutions by using off-the-shelf RGB+D cameras (e.g. Kinect or RealSense sensors) and of volumetric representations as meshes or point clouds are also possible, but they currently do not provide yet the high resolution that is required for professional and highly immersive VR content.

Moreover, there is literature comparing full 3D experiences with 360 degrees panoramas and videos in different applications. In [1], a comparison between the visualization of an archaeological site reproduced in 3D or captured in 360 pictures showed an advantage to the 3D experience in terms of presence and fun. However, we note that the implementation of the 360 picture was unusual, and the content was placed in a sphere that was stationary in space and that allowed the movement of the virtual camera in it. Moreover, Theophilus et al. [13] explored the use of live 3D reconstruction and live VR360 stream for remote collaboration in a mixed reality scenario, with an Augmented Reality (AR) user that streamed local information and a VR user that assisted the first user in a task. A comparison between both content formats actually showed better remote collaboration performance while using the VR360. However, the advantage seemed to be related to the fact that the VR360 was more competent in conveying the focus of attention of the AR user, which is crucial for carrying collaborative tasks.

*Pros of Full 3D*

– In Full 3D, all elements of the VR environment are three-dimensional. Therefore, the VR environment can be fully explored, supporting POV translation and rotation, without any extra production cost. This also means that the users can freely navigate around the 3D environment, providing 6DoF experiences.
– Full 3D is probably the most immersive content format in terms of geometric reliability and depth estimation.
– Given the absence of video components, Full 3D environments are free from compression artifacts, resulting also in a relatively lightweight option for content distribution.
– Volumetric characters can be seamlessly integrated in Full 3D environments, without needing any specific transformation. In addition, when using the case of pre-rigged 3D characters, the characters can be animated in live scenarios by just sending their data movements, which extremely reduces the transmission and processing load.
– In Full 3D, it is also possible to adapt and amend the cinematic content at a relatively low cost, or even at no extra cost. For example, different animations could be prepared and executed to respond to specific users actions (e.g., point or gaze at the user, specific answers).

*Cons of Full 3D*

– In Full 3D, the main drawback is that it is very challenging and costly — in terms of time and money — to achieve very realistic and natural photorealistic rendering and animations. This is especially true in real-time content and for 3D characters.
– In the case of pre-rigged and animated 3D avatars, a 3D scanner and a MoCap system and room need to be available.
– Meticulous post-production tasks are typically needed to refine the 3D avatars representations and animations, which have also an impact on the production costs.
– In terms of real-time volumetric users capturing solutions (e.g. by using meshes or point clouds), they still do not provide the high definition required for professional and realistic VR content, which will impact the users QoE and immersion, especially if used for the 3D actors / characters that integrate the content.

## 2.3 Hybrid 3D and video billboard

Hybrid 3D and video based solutions for VR content production is also possible (see Figure 5). The idea consists of validating if an appropriate combination, integration and blending of these content formats can contribute to leveraging the pros of the 3D and video based solutions, while overcoming their cons, at least to a certain extent. This can have an impact on the production costs, but also interestingly on the user experience.

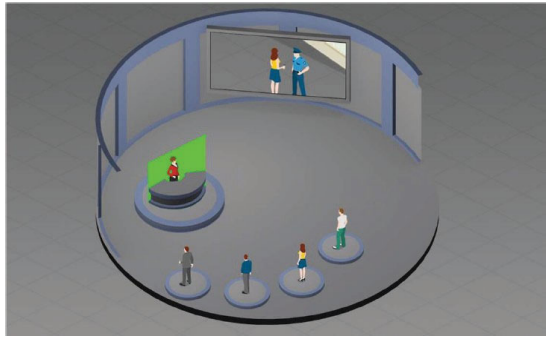

Fig. 5: Example of a Full 3D+billboard environment.

Notably, billboard impostors are used to represent background elements or objects that are too far away from the point of view to be properly identified. These were also popular on early 3D games, when graphics processing hardware was not widely available. Video billboards have also found space in the reconstruction video streams, Hayashi et al. [7] use billboards to represent football players in streams from live sports events as a 3D entity in space, independent of the field plan. The authors take advantage of the stream of video from multiple view sources to produce a billboard per camera point of view, and create a representation that is somewhat consistent from different stand points. More advanced methods [4] explore the problem of reconstructing articulated billboards based on the same kind of input video. The method reconstructs football players as an ensemble of video billboards, creating separate billboards for each limb segment of the player. In contrast, we examine the use of a video billboard to represent the main component of the content, or foreground elements, such as the actions and interactions of actors.

Figure 5 shows an example of a 3D scenario, with integrated 3D characters and/or end-users, but also with inserted video billboards for a presenter or instructor captured from a Chroma key room, and a 2D big screen (e.g. displaying TV-related content).

*Pros of 3D+billboard*

– It supports 6DoF for the 3D environment, with high geometry reliability, depth estimation, and without compression artifacts.
– The addition of inserted video billboards can increase the degree of realism, while reducing the production costs for specific, and especially dynamic, scenes.
– If the video billboards are added at strategic parts of the 3D environment, it can give the feeling that they are an intrinsic part of the three-dimensional environment, and not just an inserted video.
– The addition of inserted video billboards can provide support for POV rotation (the video could always be looking at the user) and limited translation for the 3D environment.
– It can support the addition of volumetric elements without the need to transform it to 2D, unlike in VR360, since these elements can be placed in the three-dimensional part of the VR environment.

*Cons of 3D+billboard*

– The difference between content formats may be noticeable, which may affect the user experience. Likewise, achieving a seamless integration and blending of heterogeneous content formats may be challenging in specific VR environments.
– POV translation is limited, and it can make the presence of the video billboard evident, resulting in parallax and deformation defects, and in an inconsistent 3D VR environment.
– As for VR360 video, the point of view for the billboard has to be defined before recording, and adding extra point of views considerably increases the recording effort. Likewise, the VR content cannot be easily modified or amended.

## 3 Content Production

This section reviews the whole production process for a professional VR content episode, including the pre-production, production and post-production tasks and steps. Note that the insights from the experiment of this paper will be applied to later evaluate Social VR scenarios as a new communication medium for interaction and communication between remote users [2], such as watching videos together, while apart. Accordingly, this section reports on two different, but connected, scenes about a crime investigation story. These two scenes will be used to evaluate in the future if two remote users feel together when watching the same or different, but connected, content at the same time, while interacting via audiovisual channels. However, just one of the scenes is used in the experiment presented in this paper.

### 3.1 Pre-production

#### 3.1.1 Ideation and Storyboard

After an initial analysis, it was decided to ideate a thriller-like plot revolving around a crime investigation story as the theme for the planned VR experiments. This was expected to provide both commercial relevance (by creating subsequent related episodes in the future) and validity for scientific experimentation. Being inspired by movies such as "*The Usual Suspects*", a decision was made to create an offline VR episode departing from the murder of a celebrity, in which two suspects are being interrogated and the two participants are expected to observe the interrogations, playing the role of inspectors.

A number of iterative and interactive design sessions were conducted to assess the most appropriate approach and scene for telling the story and to recreate the shared environment in which the users will virtually "meet". The options included: 1) an in-site scenario where the murder was committed and the users can freely explore and interact; 2) an interrogation inside the prison where the users can directly interact with the suspects; and 3) an interrogation behind a one-way mirror, like in classical police stations (see Figure 6). As the focus of the planned Social VR experiments was the interaction between the users, and not that much the interaction with the environment and other characters (left for the future experiments and episodes), the decision was to go with the third scenario.

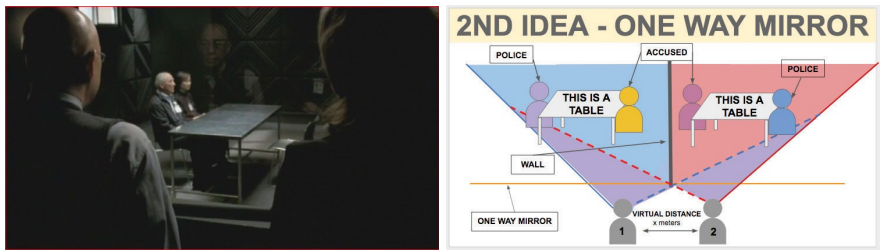

Fig. 6: Interrogation Scene through a one-way mirror.

Likewise, unlike traditional watching apart together scenarios [2], in which the users watch exactly the same content, it was decided to place the users in a shared observation room, but in front of a different one-way mirror connecting to two separate interrogation rooms (see Figure 6). In each of the separate rooms, a different suspect of the same murder is being interrogated by a policer. Therefore, although the users share a common space and can directly see and talk to each other, they can only see and hear one of the two interrogation scenes belonging to the same story. The goal was to boost the interaction and the exchange of impressions and findings between the two users in order to gather relevant hints to reach a conclusion about the authority of the crime.

Based on this concept, the storyboard including the associated spaces, viewpoints and evolution of the story was developed, in order to prepare for the production plan. As an example, Figure 7 shows the mockups that were generated to recreate the users viewpoints towards one of the interrogation scenes and the other user.

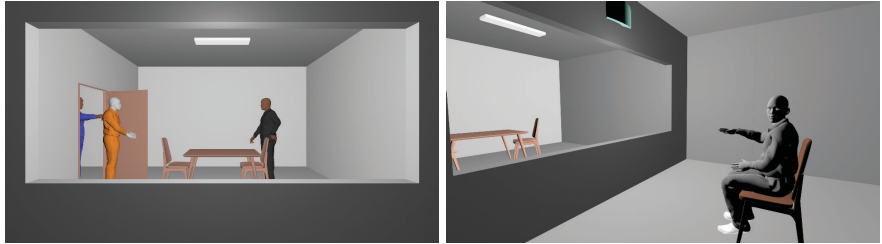

Fig. 7: users viewpoint for the produced Social VR scenario.

The experiment presented in this paper just makes use of one of the two interrogation scenes, as its goal is to determine the impact of the content formats on the user experience for single person VR experiences.

### 3.1.2 Script and Casting

After the selection of the theme and scenario, the next steps consisted of writing the script, and casting the actors. The story was further developed, revolving around the murder of Ms. Yelena Armova, a wealthy British celebrity at the peak of her career, in still unknown circumstances. Two persons are the main suspects: Mr. Ryan Zeller, the lover of the victim; and Ms. Christine Grard, her assistant. The two suspects have a different version about what happened, and the story reflects that they both have things to hide. The two suspects are being interrogated by a police inspector, Sarge. In the presented experiment, the users will experience the testimony of Ryan Zeller, but in the complete Social VR study (left for future work), the other user will see and hear the testimony of Christine Grard. Details about the casting process to select the actors (two suspects and police inspector) and the two scripts written for each of the interrogations can be found in [11]. The participation of the two actors and the actress were necessary to be able to create the three VR content formats previously introduced:

- *Full 3D* Version: A 3D environment with 3D-rigged characters, combined with MoCap techniques.
- *3D + Billboard* Version: A 3D environment with the interrogation scenes captured on video from a Chroma key room, and then rendered as a stereoscopic video billboard within the 3D environment.

– *VR360* Version: The full VR scene rendered as a stereoscopic VR360 video, composing the 3D environment and the masked video of the characters (as the 3D police station environment does not exist in reality).

In total, the VR story has a duration of 8 minutes.

## 3.2 Production

Next, the processes associated to the production of such content versions are summarized.

### 3.2.1 MoCap and shooting for the interrogation scenes

In order to create the Full 3D Version, the actors were 3D scanned with a photogrammetric scanner consisting of 96 cameras to obtain the 3D surface of their bodies (see Figure 8a). The MoCap session was recorded by using a Vicon MXT40S system with 30 cameras (see Figure 8b), in which each actor wore 59 retro-reflective markers to track their movements. For facial capture, a tool was developed to record the faces by using an iPhone X and the ARKit framework [1], with the help of an ad-hoc helmet (see Figure 8c). Then, the captured facial content data were synchronized with the MoCap data, and converted into an appropriate format for further 3D editing and adjustments.

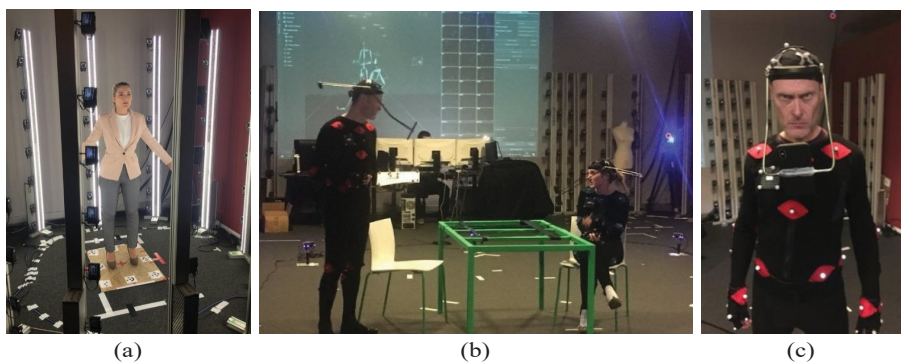

(a)                          (b)                          (c)

Fig. 8: (a) 3D body model creation, (b) MoCap recording, and (c) facial gestures recording.

In order to create the 3D + Billboard Version, the same scene, with the actors wearing exactly the same clothing, was shut over a Chroma key room, by using a stereoscopic camera (Canon, with 8-15mm optics sensors, and a separation of 8cm between its lenses). The scene objects, like the table, were also covered in green color. The recording setup can be seen in Figure 9.

---

[1] https://developer.apple.com/documentation/arkit

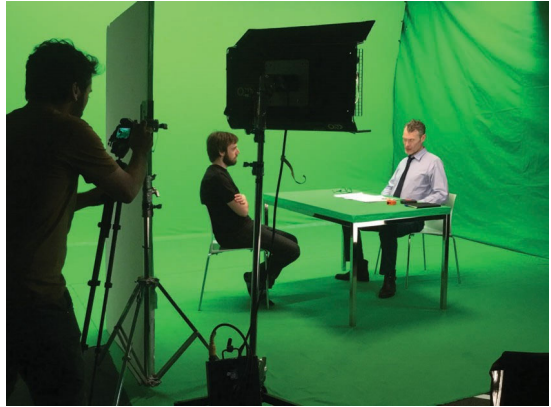

Fig. 9: Video shooting over a Chroma key room.

### 3.2.2 3D environment

The two separate interrogation rooms and the shared space for the two users, together will all associated elements (e.g. chairs, desks, book notes) were modelled in photorealistic 3D, with the use of optimized geometry, and integrated in a Unity project.

The overall view of the 3D modelled scenario views is shown in Figure 10 (left). Likewise, Figure 10 (right) shows the same 3D scenario, but with the rendering 2D video planes for the 3D + Billboard Version, together with the users positions and the shooting perspective. In both figures, it can also be observed that, for the future Social VR experiment, each user can only see (and only hear) the interrogation scene happening in front of their one-way mirror, but they can see and hear each other through a shared space. The scenes for User 1 were the ones used in the experiment presented in this work.

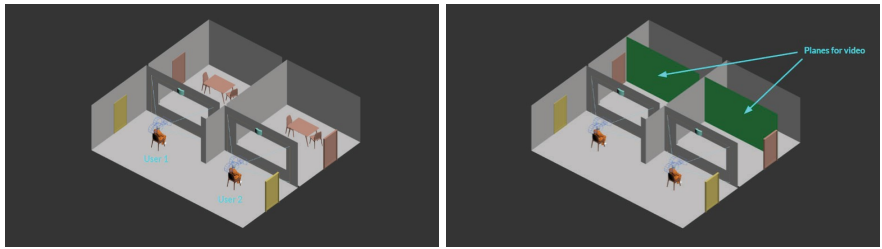

Fig. 10: (left) Overall view of the 3D environment to be recreated; (right) 3D scenarios with 2D video billboards for the interrogation scenes.

Realistic lighting conditions were recreated in order to provide a natural integration of the users and characters into the 3D virtual environment, and to provide a thriller-like atmosphere (i.e. direct light in the interrogation rooms and dimmed lights in the dark shared room where the users are placed). Spatial ambient sound was prepared, coming from the direction of the actions. This is the case for the doors opened, actions by the actors, sound from the other user, etc.

An overall view of the recreated 3D environment, resembling a 70s look police station room, is provided in Figure 11a, while the viewing perspective from one of the users through the shared space in provided in Figure 11b.

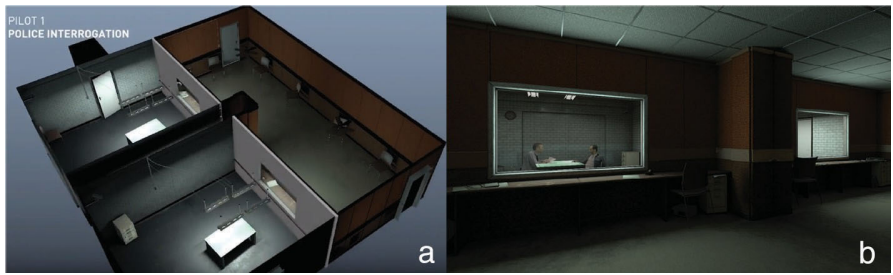

Fig. 11: (a) Overall view of the recreated 3D environment; (b) Users viewpoint from the 3D shared space.

### 3.3 Post-production

After the recording and modelling of all assets, post-production processes were conducted for all the raw material, including the associated adjustment tasks for an appropriate compositing and seamless blending.

In the case of the 3D + Billboard Version, noise reduction and masking processes were initially conducted for the recorded billboards. Masking was especially a time-consuming and laborious process, as it required the adjustment of more than 40000 frames, as well as the mask of the characters and elements of the scene. Figure 12 gives an idea of this process, including the necessary specific treatment for the characters hair. In addition, color adjustment processes were necessary for an effective removal of the green elements and the replacement with the appropriate color, together with the adjustments to achieve a seamless stereo view.

In the case of the Full 3D Version, the captured 3D surfaces of the actors were initially fit to a template character rig (i.e. humanoid skeleton with joints and bones) used to animate the characters. Then, post-processing techniques were necessary to produce morph targets that comply with the facial capture data, to clean the MoCap data (e.g. by resolving occlusions), and to retarget the animations on the character rigs to obtain realistic and natural results.

Screen captures of the final results of the produced VR content scenes, for each described variant, are shown in Figure 1. After some initial adjustments and preliminary tests, it was discovered that in the 3D + Billboard Version, a better experience in terms of motion parallax was provided if the video plane was placed slightly behind the 3D mirror with a slightly bigger size (Figure 1c). Therefore, this adjustment was applied for the experiment.

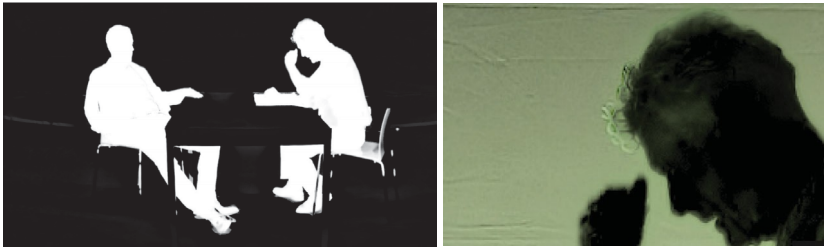

Fig. 12: Examples of the masking process.

All created content assets have been released to the Zenodo open repository [11].

### 3.4 Cost analysis

In terms of temporal costs, the Full VR360 video based Version would be the cheapest one in case the scenario exists, because everything can be directly captured with a camera. Likewise, 3D scanners can also be used for existing scenarios. In our case, the 3D scenario was non-existing and thus created from scratch. The estimation of temporal costs for the production of the 3D environment and elements is around 8 Person Months (PMs). This includes the 3D modeling and post-production tasks, like texturing, lighting, adjustments and rendering. For the 3D characters, the bodies scanning, MoCap sessions and the associated post-production and integration tasks took around 4 PMs. These processes require the availability of a 3D body scanner, a MoCap studio and a full performance capture system for body and facial capture. The production and post-production tasks for the video billboard took around 6 PMs, including the Chroma cleaning, color grading, lighting, compositing and integration. These processes require the availability of a professional stereoscopic camera and a Chroma key room. The production of the Full VR360 Version from the 3D + Billboard one just required an appropriate rendering of the scene. Then, all content versions needed to be integrated in Unity, which took around 4PM.

In terms of comparison, it can be said that the 3D + Billboard Version was a bit cheaper than the Full 3D one, in terms of required resources and time, but that was not the case for the created content, as special care was paid to the post-production tasks of the former, given that initial internal tests showed that it provided the best results for the user experience.

Note that the costs associated to pre-production tasks and to the participation of actors have not been considered, as they apply to each version. In global terms, the estimation of total costs for the production, integration and adjustments of the three content versions in Unity would be around 36PMs. This is a professional VR story and piece, in different content formats, which is intended to be presented and shown at film festivals [2].

# 4 User study

## 4.1 Methodology

### 4.1.1 Dependent variables

In the experiment, we collected information about simulation sickness, subjective sense of presence, quality of experience, comparative post-experiment feedback, and comments on each tested content condition.

The simulation sickness score was obtained using the Simulation Sickness Questionnaire (SSQ) [8]. The questionnaire was applied before and after each trial (i.e. each participant filled the questionnaire six times), the score for each trial is computed as *score after trial* minus *score before trial*, so that any abnormal state of the participant at the start of a trial is accounted for in the results. For instance, if at the start of a trial the participant presents a symptom that was acquired before the experiment or that has persisted from a previous trial of the experiment, this symptom is recorded and expected to remain until the end of the current trial, and is then subtracted from the SSQ score at the end of this trial.

The subjective sense of presence was taken after each trial, it consisted of a modified SUS (Slater, Usoh and Steed) presence questionnaire [14]. In particular, the question 5 was suppressed from the questionnaire as we considered it unsuitable in a within subject experiment design. The sense of presence is commonly described as the sense of "being there", in the virtual world, or as the feeling of non-technological mediation. That is, when the immersive equipment becomes transparent to the user.

In addition, we developed a QoE questionnaire to address four relevant aspects for the user experience, namely the quality of the virtual characters (realistic look and motion), the visual consistency of the scene (perspective projection and image composition), the feeling of control of the virtual viewpoint, and the overall experience (VR and content). The questions are presented in Table 1.

Finally, we also developed a comparison questionnaire to be applied at the end of the experiment. In this questionnaire, participants had to order the three content presentation conditions from most to least preferred with regard to the same aspects addressed in the QoE questionnaire above. The comparison questionnaire is presented, together with results, in Figure 14.

---

[2] A video describing the created VR content, and summarizing the production process, is available at: https://youtu.be/aHO5M1qNmjY

### 4.1.2 Procedure

In the user study, participants were asked to read an information sheet and sign an informed consent form. Then, they were asked to fill in a characterization questionnaire asking about their gender, height, age and previous experience with VR and video games, and were presented with an overview of the experiment structure and task. Following the introduction, participants underwent the three experimental trials, one for each content condition. The presentation order of the trials was counterbalanced to control for order effect. Each trial consisted of a pre-trial SSQ questionnaire, the cinematic segment in the current condition, a post-experience SSQ questionnaire, the adapted SUS questionnaire and the quality of experience questionnaire.

In preparation for the cinematic segment, participants were positioned sitting in a chair in the center of the capture space and equipped with the Oculus Rift head mounted display. They were also informed that they could stand up during the cinematic content if they wished to do so. This additional degree of control was permitted to help leveraging the advantages of head position tracking in the two conditions where this was allowed (*Full 3D* and *3D + Billboard*). However, participants were not allowed to walk since the tracking space was rather constrained.

Lastly, after experiencing the three content conditions, participants were asked to fill in a comparison questionnaire and to provide written feedback on each of the three content condition.

## 4.2 Participants

We recruited 24 volunteers, with a total of 8 female participants and an average age of 38 years old (standard deviation of 7.6). Five participants were using a head mounted display for the first time, while nine reported to have used it few times in the past, seven reported to use it every month or week, and three reported to use it every day. Similarly, five participants reported that this was their first VR experience, eight had few previous experiences, two participants indicated that they often used VR experiences, and nine develop a professional competence in the field of VR. The participants were all provided with a description of the experiment and had to sign an informed consent form in order to take part in the experiment.

## 4.3 Results

### 4.3.1 Simulation sickness

For each content condition, we tested whether the difference between the SSQ reported after the trial was significantly different than the SSQ reported before the trial. The statistical analysis was carried out using the Wilcoxon signed rank test. We observed a statistically significant increase in the SSQ responses for the VR360 video condition ($p = .002$). The test failed to reject the equality of pre/post trial SSQ responses in the Full 3D and 3D + Billboard conditions, which received similar SSQ scores before and after the trial ($p = .228$ and $p = .671$ respectively). Furthermore, when comparing the difference in SSQ scores across content conditions, results showed that the VR360 video condition caused statistically more discomfort than the 3D + Billboard condition ($p = .025$). An overview of the pre/post trial difference in SSQ is presented in Figure 13.

We point to two relevant factors that might be related to the increase in simulation sickness in the VR360 video condition: (1) the lack of virtual camera translation, that is, the viewpoint position is kept static even if the head of the participant translates, resulting in a visuovestibular sensory mismatch, which is known to induce discomfort and simulation sickness [9]; (2) the use of prerecorded stereo views, which can cause discomfort since the stereoscopic adjustment may not reflect the interpupillary distance of the user with precision, and cannot be adjusted by the participant.

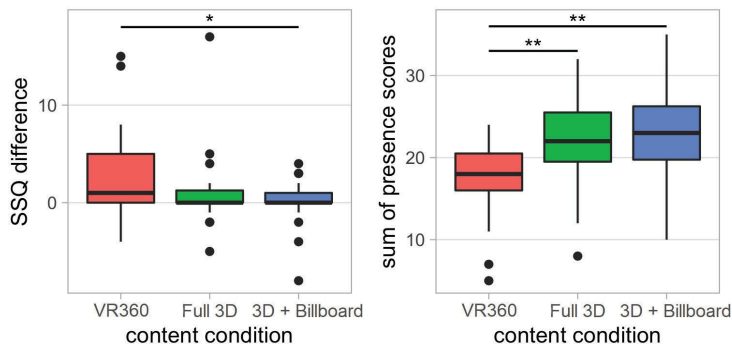

Fig. 13: (Left) Boxplot of the simulation sickness questionnaire (SSQ) score. The score was computed by subtracting the SSQ results obtained before the trial from the SSQ results obtained after the trial. The VR360 video condition presented a statistically significant increase in the SSQ, in addition, its increase was also statistically higher than for the 3D + Billboard condition. (Right) Boxplot of the subjective sense of presence questionnaire scores. Participants reported significantly higher sense of presence for the Full 3D and 3D + Billboard conditions than for the VR360 video condition. '*' and '**' indicates a significant difference with $p < .05$ and $p < .01$ respectively.

### 4.3.2 Presence

A Friedman test showed a significant effect of content condition on subjective presence score ($\chi^2_{(2)} = 12.7$, $p < .002$). Pairwise comparisons (Wilcoxon signed rank test with Holm-Bonferroni correction) across the levels of content condition showed that participants reported lower presence in the VR360 video condition than in both the 3D character ($p$ = .008) and 3D + Billboard ($p$ = .005) conditions. The latter two presented similar presence scores ($p$ = .749).

In fact, the sense of presence has been associated to accurate sensorimotor contingencies [12], that is, the coupling of motor commands and appropriate sensory feedback to how these commands affect the environment (the point of view in this case). This confirms our assumption that the VR360 video condition was going to present an inferior score in comparison to the other two content conditions because it only maps head rotation (and not translation) movements into actual virtual movement. Moreover, the insertion of video elements into a 3D virtual environment (3D + Billboard) did not seem to affect the sense of presence when compared to its full 3D counterpart. Both of these conditions afforded full control of the POV of the participant, although we should note that the tested scenario did not encourage wide movements, which could be detrimental to the *3D + Billboard* condition in terms of billboard image distortion.

### 4.3.3 QoE and comparison between conditions

For the comparative questionnaire, a chi-squared test was used for each comparison statement to determine whether a statistically significant dependence between the independent variable condition (Full 3D, 3D + Billboard or VR360 video) and the dependent variable classification (1st, 2nd and 3rd – or best, intermediate, worst), as specified by participants, exists. All but one of the tests showed a statistically significant dependence between the variables (all $p$ < .001). The statistical test failed to reject the independence between the variables for the statement "the visual consistency between characters and scenario was more accurate in condition ..." ($p$ = .112). A summary of the comparison questionnaire results is presented in Figure 14.

With regard to the QoE questionnaire, we tested each response variable for statistically significant differences across the levels of content condition using the Friedman test. Significant results were followed by pairwise comparisons using the Wilcoxon signed rank test and Holm-Bonferroni correction. A summary of results is presented in Figure 15. We discuss each of the four addressed QoE aspects in the questionnaires below.

The questions concerning expressive motion, natural motion, and realistic appearance of the virtual characters (Q1, Q2 and Q3 respectively) presented similar results, with 3D + Billboard and VR360 characters being considered to have more realistic appearance and more natural and expressive movement than Full 3D (all $p$ < .01). In the post-experiment comparative questionnaire, the 3D + Billboard condition received higher scores on character appearance and animation, followed by VR360. These results indicate the superiority of recorded video media when it comes to visual representation of character appearance and actions. Participants observed detrimental animation effects in the Full 3D condition, such as unrealistic hand movements and the lack of physical simulation on characters clothing and props.

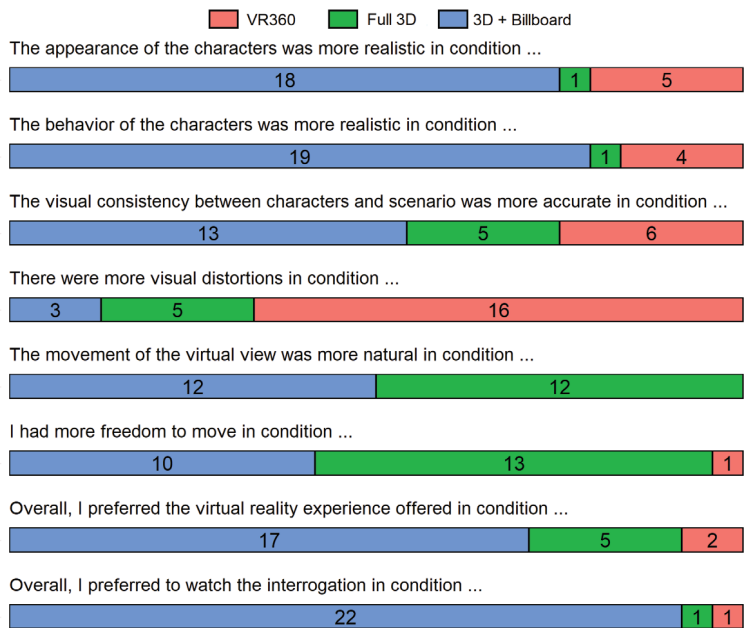

Fig. 14: Comparison questionnaire results. Participants classified each condition from most (1) to least (3) preferred for eight different statements.

Considering the perception of correctness of visual perspective (Q4), movement of virtual objects through space (Q5) and visual consistency of scene elements (Q6), the 3D + Billboard condition performed better than the VR360 video condition (all $p < .01$), while Full 3D performed significantly better than VR360 video considering visual perspective and scene elements (Q4 and Q6, both $p < .01$), but not with regard to the movement of virtual objects (Q5, $p = .34$). We did not find a statistically significant difference in Q4, Q5 and Q6 when comparing the 3D + Billboard and Full 3D conditions. The post-experiment comparative questionnaire presented similar results. The VR360 video condition was argued to have more visual distortions, while the Full 3D and 3D + Billboard conditions received similar scores.

Moreover, the feeling of control (Q7 and Q8) was stronger on the 3D + Billboard and Full 3D conditions than on the VR360 condition (both $p < .001$), while no significant difference was found between the 3D + Billboard and Full 3D conditions (both $p > .4$). This outcome confirms our assumption, given that the VR360 was the only condition not to allow for POV position updates in the virtual environment. This difference was not observed when comparing the 3D + Billboard and Full 3D conditions. The comparison questionnaire produced similar results. The movement of the virtual viewpoint was considered the most natural in the Full 3D or 3D + Billboard conditions by nearly the same amount of participants, while VR360 was generally considered worse. Similar results were also observed for participants perception of freedom to move, but with a slight advantage for the Full 3D condition over the 3D + Billboard condition.

Table 1: QoE questionnaire, answers were provided in a 5 point scale ranging from strongly disagree (1) to strongly agree (5).

| ID | Question | Aspect of QoE |
|---|---|---|
| Q1 | The motion of the characters felt expressive. | Characters |
| Q2 | The motion of the characters felt natural. | Characters |
| Q3 | The appearance of the characters is realistic. | Characters |
| Q4 | The perspective of the virtual world and the elements in it looked consistent. | Visual consistency |
| Q5 | The movement of virtual elements, character, objects ... through the space felt consistent with the real world. | Visual consistency |
| Q6 | My movement did not affect the visual consistency of the virtual world and the elements in it. | Visual consistency |
| Q7 | The control of the virtual view felt natural. | Feeling of control |
| Q8 | When I moved, I expected the virtual view to respond accordingly. | Feeling of control |
| Q9 | I felt involved in the virtual reality experience. | Overall experience |
| Q10 | My experiences in the virtual environment seemed consistent with my real world experiences. | Overall experience |
| Q11 | I felt involved in the interrogation. | Overall experience |

Lastly, the 3D + Billboard and Full 3D conditions were ranked higher than VR360 video condition for overall VR experience and experience consistency with the real word (Q9 and Q10, both $p < .01$). Concerning the interrogation experience (Q11), the 3D + Billboard condition ranked higher than the VR360 video condition ($p = .036$), while Full 3D presented a score that could not be differentiated from the other two conditions (both $p = .32$). For all three questions, no statistically significant difference was found between the 3D + Billboard and Full 3D conditions. In the comparison questionnaire, the 3D + Billboard condition was generally the preferred condition for this VR experience. It was then followed by the Full 3D condition, in spite of the perception that the virtual characters appearance and behavior were inferior in this particular condition.

Notably, the overall experience scores for the Full 3D condition were similar to the 3D + Billboard condition scores and generally superior to the scores given to the VR360 video condition. Our results indicate that, in spite of presenting non photorealistic content (Q1, Q2 and Q3), participants held a positive view of the Full 3D condition in terms of the overall experience. In addition, we observed an overall preference for the 3D + Billboard condition when the direct comparison between the three conditions was possible. This indicates that video billboards are a suitable form of content representation in VR experiences.

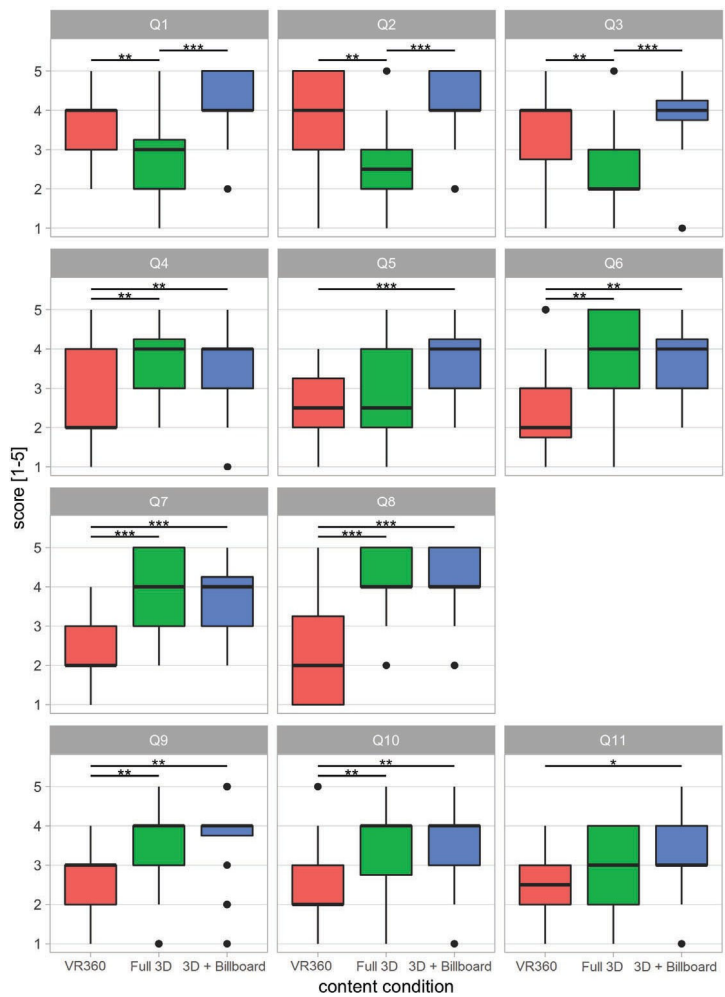

Fig. 15: Boxplots of the quality of experience questionnaire. '*', '**' and '***' indicate a statistically significant difference with $p < .05$, $p < .01$ and $p < .001$ respectively.

### 4.3.4 Participants comments

At the end of the experiment, participants were asked to write down the advantages and disadvantages of each of the three content formats, as assessed by themselves. The objective was to identify the most significant and salient characteristics of each experience from the perspective of participants. The feedback was collected as a short statement in a digital form. Observations with equivalent or closely related statements were grouped together. We report on the characteristics that were cited more often for each of the conditions.

The advantages reported for the VR360 video condition were the realism of characters (4 participants) and environment (6 participants) and the consistent quality in terms of character/environment integration (2 participants). However, 7 participants reported that they did not see any advantage in the VR360 video condition. The main disadvantages were the lack of position tracking (12 participants), which led to simulation sickness and is supported by the SSQ results (3 participants), the presence of compression artifacts (6 participants), and decreased immersion (3 participants).

The advantages reported for the 3D + Billboard condition were the realism of characters and environment (14 participants), the comfort and/or freedom to move (8 participants) and increased immersion (7 participants). The disadvantages were the visual inconsistency between the 3D environment and the billboard video (4 participants), which was particularly noticeable in the table in the interrogation room, the resolution of the video (2 participants), and the flatness of the billboard (2 participants), which felt like a screen. Notably, 8 participants did not express any disadvantage for the 3D + Billboard condition.

Finally, the advantages reported for the Full 3D condition were the consistent visual experience (8 participants), the freedom to move (6 participants), realism of the 3D environment as the adjacent room is not longer contained in an image (4 participants) and increased immersion (4 participants). In addition, 3 participants did not see any advantage for this condition. The most common disadvantages were the characters appearance (15 participants) and acting (12 participants), which did not look as realistic as in the other two conditions.

## 5 Conclusions

In this paper, we have presented a discussion and experiment on the subject of VR content production and consumption. For such a purpose, a professional VR content episode has been produced, using three content format variants, as stimuli for the experiment. In the experiment, participants tested three different content conditions, a VR360 video, a combination of video and 3D environment (3D + Billboard), and a Full 3D condition. Under our experimental conditions, participants were generally more receptive to the content condition that combines video content and 3D environment, despite the fact that this particular condition often produces incorrect perspective projection in response to changes in the point of view. In fact, the addition of motion parallax (POV translation) while preserving the visual quality of the video content was perceived as a major advantage of the 3D + Billboard condition.

Overall, our results show that most participants had the best experience in the 3D + Billboard condition. The 3D + Billboard condition presented subjective presence scores that are similar to the Full 3D condition and higher than the VR360 video, very little variation in simulation sickness due to the VR experience, and was generally considered the better option on the post experiment questionnaires.

However, we should point out a few limitations concerning the design of our experiment and the generality of our results. Notably, our experiment places the user and the content in different rooms. This has two main implications. First, the visual integration of the billboard video with the 3D environment becomes easier, since the billboard is only partially visible through the window opening that connects both rooms. Second, it limits the potential perspective distortion caused by video billboard/3D projection mismatch, since the user can only translate relative to the video billboard by a certain amount, and never get to the point where the billboard is seen as a completely flat object (i.e. when the image projection plane is perpendicular to the billboard plane).

Despite these limitations, we believe our work is still relevant for many VR scenarios, for which 6DoF does not become a key requirement. Noteworthy examples are Social VR scenarios [10, 5, 2], like e-learning or multi-user conferencing, or shared video watching.

To address and gain deeper insight into these limitations, future work will focus on investigating the impact of different combinations of content formats in VR scenarios with 6DoF capabilities, and enabling different forms of interaction.